\documentclass[aps,pra,twocolumn,showpacs,superscriptaddress,amsmath,amssymb]{revtex4}
\usepackage{graphicx}
\usepackage{dcolumn}
\usepackage{bm}
\usepackage{amsmath}

\newcommand{\bra}[1]{\langle #1|}
\newcommand{\ket}[1]{|#1\rangle}

\begin{document}

\title{Analytic models of ultra-cold atomic collisions at negative energies for application to confinement-induced resonances} 
\author{S. G. Bhongale} \email{bhongale@rice.edu}
\affiliation{Department of Physics and Astronomy, MS-61 Rice
  University, 6100 Main St., Houston, TX 77005, USA.}
\affiliation{Department of Physics and Astronomy, University of New
  Mexico, Albuquerque, NM 87131, USA.}
\author{S. J. J. M. F. Kokkelmans} \affiliation{Eindhoven University
  of Technology, P. O. Box 513, 5600MB Eindhoven, The Netherlands.}
\author{Ivan H. Deutsch} \affiliation{Department of Physics and
  Astronomy, University of New Mexico, Albuquerque, NM 87131, USA.}

\begin{abstract}
We construct simple analytic models of the $S$-matrix, accounting for
both scattering resonances and smooth background contributions for
collisions that occur below the s-wave threshold.  Such models are important for
studying confinement-induced resonances such as those occurring in cold collisions
of $^{133}$Cs atoms in separated sites of a polarization-gradient optical lattice.  
Because these resonances occur at negative energy with respect to the $s$-wave 
threshold, they cannot be studied easily using direct numerical solutions of the
Schr\"{o}dinger equation.  Using our analytic model, we extend previous studies 
of negative-energy scattering to the multichannel case, accounting for the interplay
of Feshbach resonances, large background scattering lengths, and inelastic processes.

\end{abstract}
\date{\today}
\maketitle

\section{Introduction}
The ability to control ultra-cold atom-atom interactions has opened the
door to a wide variety of fundamental and applied studies, including
the production of ultra-cold molecules
\cite{jin1,ketterle1,grimm1,hulet1}, simulations of condensed matter
phenomena \cite{condmatsim,quantumhall}, and quantum information
processing \cite{uwe}.  The tools that have been central to this
development include designer atom traps, for example, optical lattices
\cite{opticallattice}, and controllable scattering resonances such as
a magnetic Feshbach resonance \cite{feshbach}. Both of these can be
used to manipulate the two-body scattering process, thus affecting the
strength of the interaction, the nature of the resulting two-body
states, and more general many-body phenomena.  Examples include
confinement-induced resonances \cite{cir}, bound-states with repulsive
interactions \cite{boundneg}, and Feshbach resonances in band
structures \cite{feshbachlattice}.
 
\begin{figure}[hb]
  \includegraphics[scale=.50]{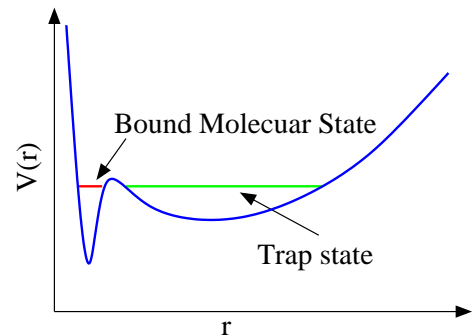}
\caption{Schematic of the effective potential between atoms trapped in
  separated wells of an optical lattice, as a function of the relative
  coordinate in the direction of trap separation (not to scale).  At
  short range there is a molecular binding potential.  At long range
  the relative coordinate is bound by the traps.  For a given
  separation and well depth, one of the trap vibrational states may
  become resonant with a bound state of the two-body interaction
  potential, resulting in a trap-induced resonance.}
\label{tirschematic}
\end{figure}

A particular example that we have explored previously is a
trap-induced resonance (TIR) that occurs as a result of interaction
between atoms that are confined to spatially separated harmonic traps
\cite{stockprl,tir}.  This happens as a consequence of a molecular
bound state that becomes resonant with the vibrational state of the
separated atoms due to a quadratic rise in the light shift when the
two atoms approach one another, as shown schematically in
Fig.~\ref{tirschematic}. A strong resonance can occur when the
confinement of the wave packet in the trap is on the order of (or
smaller than) the free-space scattering length. This $s$-wave
resonance is analogous to a higher-partial wave shape resonance
occurring in free space, but here the tunneling barrier arises from
the trap rather than from an angular momentum centrifugal barrier.
Because of this tunneling, the interaction occurs at ``negative
energy" values with respect to the free-particle $s$-wave scattering
threshold.

In previous work on the TIR in cesium \cite{tir}, whose large
scattering length gives rise to a strong resonance, Stock {\em et al.}
extracted the scattering length at negative energy for the single
channel case by an explicit integration of the radial Schr\"{o}dinger
equation at negative energies using the Numerov method. Such a
procedure has limited utility; the solutions are unstable since the
wave function blows up in the tunneling barrier. The situation gets
very complicated as soon as there is more than one channel. A proper
numerical technique must ensure that open channels are propagated
along with the exponentially decaying closed channels while
maintaining accuracy to relevant digits. Most coupled channel codes
that incorporate such situations (i.e. propagating closed and open
channels), eventually drop the closed channels beyond certain radius
since they are only interested in open channels.  On the other hand,
the problem of TIR considered in this paper requires us to integrate
to large enough radius for determining a good asymptotic log
derivative (for both open and closed channels). We emphasize that we
are not allowed to drop any channels, since in the end, we are
required to extract the asymptotic log derivatives for both open as
well as closed channels.

To remedy this, we consider here {\em analytic models} of the multichannel
$S$-matrix.  Simplified analytic models have been employed in previous
studies of ultra-cold collisions and scattering resonances. Julienne
and Gao have predicted the properties of Feshbach resonances based on
the analytic properties of the van der Waals long range potential
\cite{julienneandgao}.  Marcelis {\em et al.} have used analytic
models to describe the interplay of open and closed channels in the
context of Feshbach resonances associated with a large background
scattering length \cite{marcelis}. These analytic models, while simple
in nature, are able to encapsulate the necessary physics in just a few
parameters. These parameters can then be incorporated into building
model many-body Hamiltonians, an example being the
two-channel model used for describing resonance superfluidity in a two
component Fermi gas \cite{kokkelmansandholland}.

Our goal in this article is to develop an analytic model that can be
used to study the TIR for Cs atoms trapped in an optical lattice. In
Sec. II we review the basic physics that gives rise to the TIR, its
application in Cs, and show the limitations of direct numerical
solutions, even for single channel scattering.  Section III contains
the heart of our new results.  We review the basic resonant scattering
phenomena and how they are modeled analytically in the $S$-matrix.  We
then apply this to determine expressions for the negative-energy
scattering length in a nontrivial multichannel scattering process,
relevant to an experimental observation of the TIR.  We summarize our
results in Sec. IV.

\begin{figure}[t]
  \includegraphics[scale=.45]{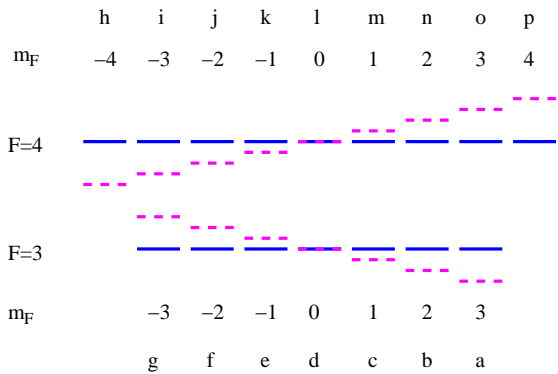}
\caption{Energy level diagram of the 6S$_{1/2}$ hyperfine manifold of
  $^{133}$Cs (not to scale). The solid (dashed) lines correspond to the levels in the
  absence(presence) of the an external magnetic field. The magnetic
  quantum numbers and the corresponding level-labellings are shown.}
\label{levels}
\end{figure}

\section{Scattering resonances in $^{133}$Cs}\label{resonances}
We consider the scattering of two $^{133}$Cs atoms in their 6S$_{1/2}$
electronic ground state, trapped in an optical lattice.  The Zeeman
hyperfine structure of this manifold is shown in the Fig.~\ref{levels}
with magnetic sub-levels labeled as $a, b,\ldots p$ for
convenience. Henceforth, all two-atom scattering channels will be
denoted by the relevant pair of these sub-levels.  To begin with, we
consider the scattering in the $\ket{ap}$ channel. This is motivated
by studies of controlled collisions via spin-dependent transport in
polarization gradient lattices \cite{polgradcollisions}.  In these spin
states, two atoms that are separated by $\lambda/4$ in a lin-perp-lin
optical lattice can be transported into the same well in a
lin-parallel-lin optical lattice via a rotation of the laser
polarization. By angular momentum conservation, because these are
``stretched states", and ignoring small spin-dipolar and second-order
spin-orbit coupling \cite{leoandwilliams}, the dominant $s$-wave
collision does not couple this channel to any other channel.  The
result is an elastic phase shift that can be used to implement an
entangling two-qubit logic gate. In addition, as the wells approach
one another, there will be a TIR that can strongly affect the two-atom
interaction \cite{stockprl}.

\begin{figure}[b]
  \includegraphics[scale=.45]{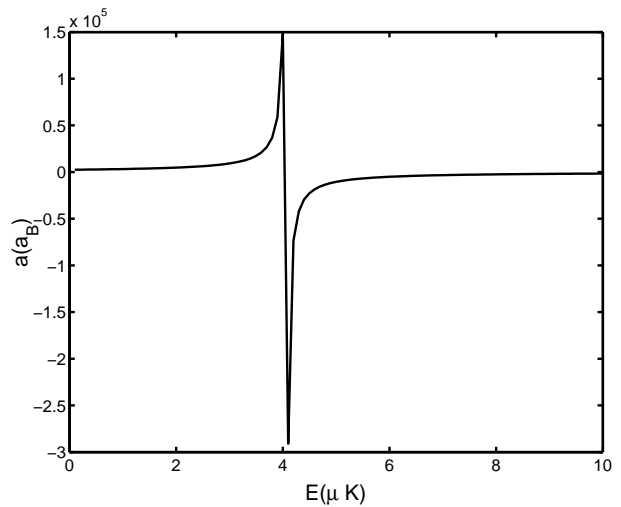}
\caption{Scattering length for the $\ket{ap}$ channel in units of Bohr
  radius $a_B$ as a function of scattering energy $E$.}
\label{rvalues}
\end{figure}

The properties of the TIR follows from a simple model of the two-atom
system.  We express the Hamiltonian for the atoms in the
$\ket{ap}$ scattering channel in center-of-mass coordinates,
$\mathbf{R}$, and relative coordinates, $\mathbf{r}$, as 
\begin{eqnarray}
H_{\text{CM}}&=&\frac{P_R^2}{2M}+
\frac{1}{2}M\omega^2R^2,\\ 
H_{\text{rel}}&=&\frac{p^2}{2\mu}+\frac{1}{2}\mu\omega^2 | \mathbf{r}-\Delta
\mathbf{z}|^2+V(r)\label{hrelative},
\end{eqnarray}
where $\mu$ is the reduced mass, $\Delta \mathbf{z}$ is the separation
of the traps, and $V(r)$ is the inter-atomic potential. In principle,
the TIR can be seen by diagonalizing the above Schr\"{o}dinger
equation using the precise Cs$_2$ molecular potential projected on
$\ket{ap}$ for $V(r)$. This is a non-trivial task, however, since
there is a huge separation of length scales between the molecular
potential and the external trapping potential, and the displacement of
the trap from the zero of the relative-coordinate makes the system
anisotropic.  Instead, we treat the molecular potential through a
contact pseudo-potential \cite{bolda},
\begin{equation}
V(r,E)=\frac{2\pi\hbar^2}{\mu}a(E)\delta(r)\frac{\partial}{\partial r}.
\end{equation}
Here $a(E)$ is the energy-dependent $s$-wave scattering length,
determined by direct numerical integration of the Schr\"{o}dinger
equation based on the $s$-wave scattering phase shift of the known
Cs$_2$ molecular potential in the {\em absence} of the trap, according
to $a(E)=-\tan \delta_0 (E)/k$. The energy is then chosen
self-consistently to solve the Schr\"{o}dinger equation, including
both the boundary conditions at short-range due to the atomic
interaction and at long-range due to the trap \cite{stockprl}.  For
sufficient separation between the traps, the lowest energy eigenstates
drops below the threshold of the molecular potential.  Thus the
``negative energy" scattering states that are inaccessible in free
space, become opened by the trapping potential.

The scattering length at positive energies for the $\ket{ap}$ channel,
calculated using a numerical solution to the radial Schr\"{o}dinger
equation based on the well-established $^{133}$Cs dimer potential, is
shown in Fig.~\ref{rvalues}.  A resonance exists at $E=4.03$ $\mu$K
(here and throughout, energy is measured in temperature units) due to
a bound state very close to zero energy.  Finding the scattering
length at negative energies via equivalent numerical integration is
highly unstable, as the wave function blows up in the classically
forbidden region.  Even if one manages to do it, it is necessary that
the integration be sufficiently stable for large $r$, well into the
asymptotic region of the potential, in order to extract a meaningful
scattering length \cite{tir}. In addition, because we are in the
neighborhood of a scattering resonance near zero energy, the strong
variation of the scattering length with energy makes the requirement
for a robust numerical solution even more demanding.  To address these
problems, we develop analytic models that will allow us to calculate
the scattering matrix below threshold. In doing so, we will extend the
method to a more complicated process that occurs for the multichannel
scattering case.

\section{Analytic model}\label{model}
Near-threshold scattering is dominated by resonant phenomena.  Such
resonances can arise from a variety of different physical mechanisms,
and thereby affect the form of the analytic model.  We identify the
nature of the resonance based on our understanding of the physical
processes and the location of the poles in the $S$-matrix.  Away from
resonance, the scattering properties are smooth functions and
therefore can be modeled by a few free parameters.  The total
$S$-matrix thus factors into resonant and background contributions,
\begin{equation}
S(k)=S_{\text{bg}}(k)S_{\text{res}}(k).
\end{equation}

\begin{figure}[t]
  \includegraphics[scale=.45]{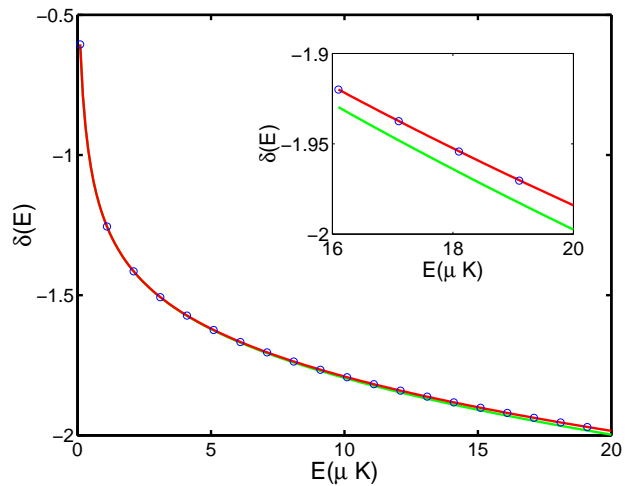}
\caption{Energy dependent phase shift $\delta(E)$ for $s$-wave
  scattering on the $\ket{ap}$ channel. The circles represent numerical data from coupled
  channel calculation, the green curve is the analytical fit using
  Eq.~(\ref{linearphase}), and the red curve is the analytical fit
  using Eq.~(\ref{effectiverange}).}
\label{phasefit}
\end{figure}

\begin{figure}[b]
  \includegraphics[scale=.45]{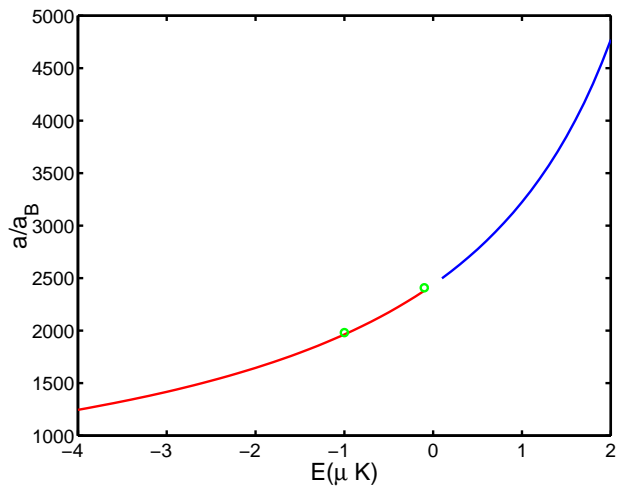}
\caption{Analytic continuation of the scattering length for the
  $\ket{ap}$ to negative energies.The circles represent data from
  numerical integration of the Schr\"{o}dinger equation.}
\label{scatfit}
\end{figure}

For the scattering on the single channel $\ket{ap}$, the resonant
behaviour can arise only from a bound state or a virtual-bound state
near threshold; any other type of resonance such as a Feshbach will
require the inclusion of other channels.  A direct numerical
integration gives a bound state with a binding energy $E_b=246.2$
$nK$. Thus, the resonant part of the single channel $S$-matrix can be
written as \cite{jochain},
\begin{eqnarray}
S_{\rm{res}}(k)&=&-\frac{k+i\kappa_{\text{b}}}{k-i\kappa_{\text{b}}},
\end{eqnarray}
where $\hbar^2\kappa_{\text{b}}^2/2\mu=E_b$, representing
a pole of the $S$-matrix in the $k$-plane on the positive
imaginary axis. There is no need to single out the other bound
states, as they are energetically too far away, and their effect is absorbed
in the background part. This remaining part can be written in the low energy
limit $k\rightarrow 0$ as,
\begin{equation}
S_{\text{bg}}(k)=\exp(-2ia_{\text{bg}}k),
\end{equation}
where $a_{\text{bg}}$ is the background scattering length that
encapsulates the effect of all other non-resonant processes, including
other deeply bound states. Now we can write the complete $S$-matrix
element analytically with just one free parameter, $a_{\text{bg}}$,
the value of which can be determined by fitting one positive low
energy point to the equation
\begin{equation}
\delta(k) = -a_{\rm{bg}}k-\tan^{-1}(k/\kappa_{\text{b}}).\label{linearphase}
\end{equation}
 In Fig.~\ref{phasefit} we plot the scattering phase shift as a
 function of the scattering energy, obtained via the full coupled
 channels calculation and compare it with the one obtained from
 Eq.~(\ref{linearphase}). We see good agreement at low energy but for
 energies beyond 10 $\mu$K a slight difference is noticed. This is
 expected since the linear form of the background phase shift is only
 valid at low energies. To remedy this problem, we use a higher order
 expansion for the background part given by
\begin{equation}
S_{\rm{bg}}(E)=\frac{-1/a_{\rm{bg}}+r_0k^2/2+ik}{-1/a_{\rm{bg}}+r_0k^2/2-ik},
\label{effectiverange}
\end{equation}
where we have added an additional free parameter, $r_0$, the effective
range.  As before, we determine both the parameters by fitting two low
energy data points. The inset of Fig.~\ref{phasefit} shows excellent
agreement of this improved model with the numerical solution.

Given the form of the S-matrix, we can predict the scattering
properties in the $\ket{ap}$ channel at negative energy values by
performing an analytic continuation of the $S$-matrix to the imaginary
$k$-axis. From this one can consistently define the scattering length
at negative energies $E=-\hbar^2\kappa^2/2\mu$ by
\begin{equation}
a(i\kappa)=-\frac{\tan(\delta(i\kappa))}{i\kappa}.
\end{equation} 
In Fig.  \ref{scatfit} we plot the scattering lengths obtained from
the analytic procedure discussed above.  These agree with the direct
numerical integration just below threshold \cite{tir}.

We now turn to a more complex situation: collisions between $\ket{a}
=\ket{F=3,m_F=3}$ and $\ket{o} =\ket{F=4,m_F=3}$ or the $\ket{ao}$
channel. This is motivated by the following experimental
considerations.  Controlled collisions via spin-dependent transport in
polarization gradient optical lattices \cite{polgradcollisions} is
hampered by inhomogeneous broadening arising from unwanted real or
fictitious magnetic fields (due to elliptically polarized light at the
atomic position) \cite{deutschandjessen}.  This is a particularly
deleterious effect for the $\ket{ap}$ states that see a strongly varying
difference in their optical potentials along the transport.  In a lin-angle-lin optical
lattice at very large detunings, atoms in the $\ket{a}$ and $\ket{o}$ states experience almost
the same shift due to the fictitious magnetic field.  Any residual
broadening is due to the finite detuning effects (giving a
differential scalar light shift) and the true magnetic field
inhomogeneity.  Controlled collisions in the $\ket{ao}$ channel thus
offer the advantage of higher degrees of coherence, with potential
applications to quantum information processing.

\begin{figure}[t]
  \includegraphics[scale=.45]{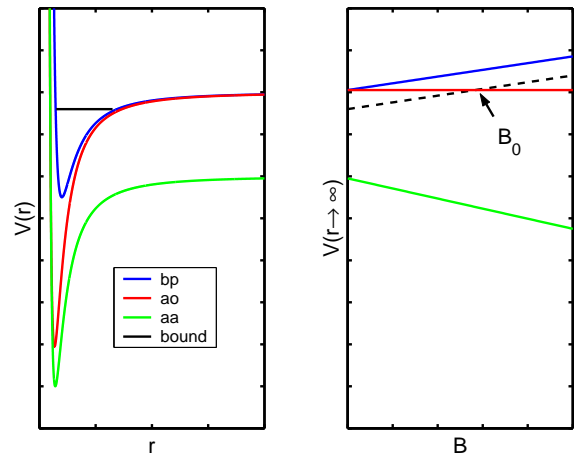}
\caption{Schematic of the potential energy curves for different
channels. The figure on the left shows potential energy as a function of
the internuclear separation. The right figure shows the scattering
threshold for different channels as a function of the external
magnetic field $B$. At $B=B_0$, the bound state corresponding to the
$\ket{bp}$ channel crosses the $\ket{ao}$ channel threshold. 
The $\ket{bp}$ channel becomes closed for $B>0$.}
\label{potschematic}
\end{figure}

To treat scattering with the incoming $\ket{ao}$ channel, we must
account for the exchange interaction, which leads to spin-changing
collisions that preserve the total projection of angular momentum
along a quantization axis.  In this case, the $s$-wave collisions
couple the $\ket{ao}$ channel to the $\ket{bp}$, $\ket{aa}$,
$\ket{oo}$, and $\ket{pn}$ channels.  At low energies, small compared
to the hyperfine splitting, $\ket{oo}$ and $\ket{pn}$ are
energetically closed.  Moreover, in the presence of any positive
magnetic field $B>0$, the channel $\ket{bp}$ shifts to a higher energy
compared to the $\ket{ao}$ channel, as depicted in
Fig.~\ref{potschematic}. For energies that are smaller than this
shift, $\ket{bp}$ channel is also closed.  The movement of this
channel from below to above the threshold can lead to a Feshbach
resonance that strongly affects the scattering process, as discussed
below.

\begin{figure}[b]
  \includegraphics[scale=.37]{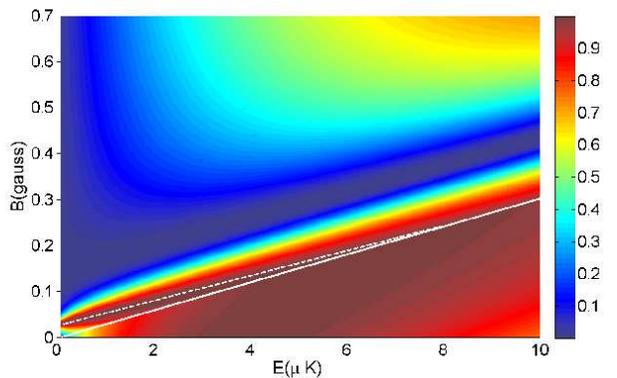}
\caption{Scattering phase for the $|ao\rangle$ channel as a function
  of energy and magnetic field, plotted as surface plot of
  $\sin^2(\delta(E,B))$.  The solid line corresponds to the boundary
  in the (E,B) plane that separates the region where channel
  $|bp\rangle$ is closed and open. The dashed line corresponds to the
  points where $\delta(E,B)=\pi/2$. The color scheme used is linear
  with dark red corresponding to the value $1$ and dark blue to $0$.}
\label{aophase}
\end{figure}

\begin{figure}[t]
  \includegraphics[scale=.42]{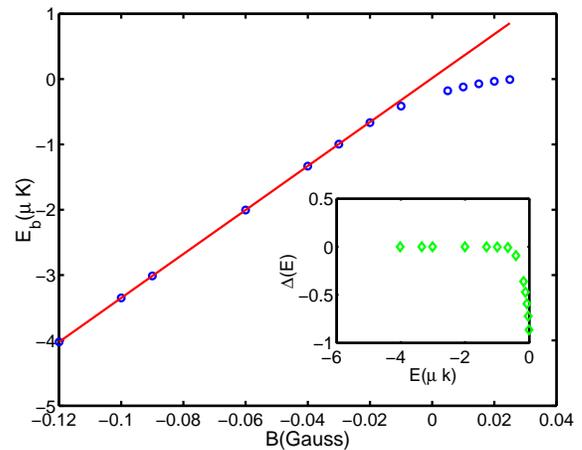}
\caption{Location of a molecular bound state as a function of magnetic
  field. Data points shown with circles correspond to values obtained
  from direct diagonalization of the two-atom Schr\"{o}dinger equation
  in a large quantization volume. The solid line indicates the
  position of the bare $\ket{bp}$ channel bound state which moves up
  with B-field. Inset shows the shift of the bound state due to
  dressing by the $\ket{ao}$ channel}
\label{boundstate}
\end{figure}

\begin{figure}[t]
  \includegraphics[scale=.53]{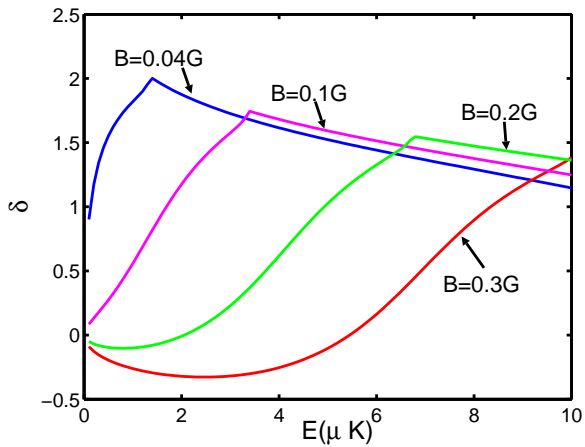}
\caption{Scattering phase shift for the $\ket{ao}$ channel as found
  from a full multichannel calculation, as a function of energy for
  different values of magnetic field B. The crossing of the
  $|bp\rangle$ channel threshold is marked by a finite jump in the
  value of $d\delta/dE$. }
\label{jumpinphase}
\end{figure}

The scattering phase shift for the $\ket{ao}$ channel is calculated by
a full coupled-channels calculation. In Fig.~\ref{aophase} we plot
$\sin^2(\delta(E,B))$ as a function of scattering energy $E$ and the
external magnetic field $B$ in the range of a few hundred mG.  As is
seen in this figure, there is a scattering resonance along the dashed
line where $\delta(E,B)=\pi/2$. Also, since the resonance moves
monotonically upwards in energy as a function of the $B$ field, it is
clear that this resonance is a Feshbach resonance. We confirm this by
calculating the bound state energy and find that it changes sign at
approximately $B=30$ mG as shown in Fig.~\ref{boundstate}. There also
exists a threshold $E_T(B)$ corresponding to the opening of the
$\ket{bp}$ channel (shown by the white solid line in
Fig.~\ref{aophase}). Across this threshold the number of open channels
changes by one, as reflected in the abrupt change in the $(E,B)$
dependence of the scattering properties. This is indicated in
Fig.~\ref{jumpinphase} where we plot the scattering phase shift as a
function of energy for various values of the $B$ field.

Based on the above understanding of the mechanisms that lead to scattering 
resonances, our goal is to build an analytic model that will allow us
to predict the scattering lengths at negative energies where the TIR
is predicted to occur.  To begin with, we will again assume that the
diagonal element of the $S$ matrix can be modeled as
\begin{eqnarray}
S_{ao}=\bra{ao} S(B,k) \ket{ao} =S_{\rm{bg}}(B,k)S_{\rm{fesh}}(B,k)
\end{eqnarray}
where $S_{\rm{bg}}$ is a smooth function describing the background
contribution and $S_{\rm{fesh}}$ is the contribution arising from the
Feshbach resonance. As before, the resonance leads to a pole in the
$S$-matrix. In fact it can be shown that every element of the
multichannel $S$-matrix has a pole corresponding to this resonance
\cite{taylor}. Also, since the location of the resonance moves upwards
in energy almost linearly as a function of the $B$ field, it is fair
to assume that the bound state solely resides on the $\ket{bp}$
channel and is shifted by an amount $\Delta_{\text{fesh}}$ due to
coupling to the $\ket{ao}$ channel.  Therefore without considering the
off-diagonal elements of the $S$-matrix, the resonant part can be
modeled by a pure Breit-Wigner pole,
\begin{equation}
S_{\rm{fesh}}(B,E) =
\frac{E-E_b(B)-\Delta_{\rm{fesh}}-i\Gamma(E)/2}{E-E_b(B)-\Delta_{\rm{fesh}}+i\Gamma(E)/2},\label{feshbachmodel}
\end{equation} 
so that that Feshbach contribution to the phase shift is
\begin{equation}
\delta_{\rm{fesh}}(B,E) =
-\tan^{-1}\left[\frac{\Gamma(E)/2}{E-E_b(B)-\Delta_{\text{fesh}}}
  \right].\label{feshphase}
\end{equation}
Here $E_b(B)$ is the location of the bare bound state of the
$\ket{bp}$ channel, $\Gamma(E)$ is the width of the Feshbach resonance
and represents the location of the pole of $S(E,B)$ along the
imaginary axis in the complex $E$ plane.  This model is restricted to
energies below the $\ket{bp}$ threshold for a finite magnetic field,
$E_T(B)$.  The discontinuity in the scattering phase shift seen in
Fig.~\ref{jumpinphase} shows that a different model is required as
additional scattering channels are opened. We leave that problem for
future study.

\begin{figure}[t]
  \includegraphics[scale=.55]{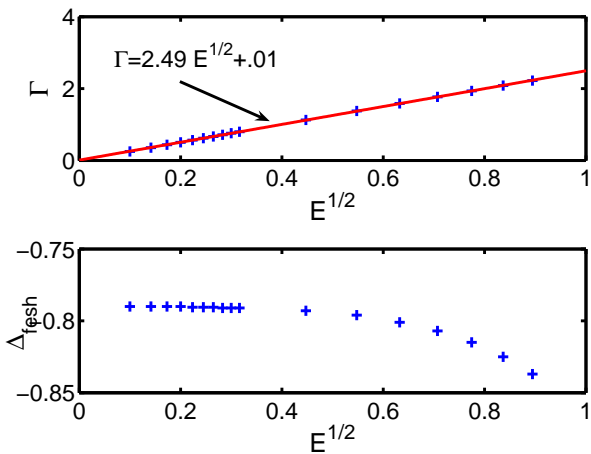}
\caption{ Top figure shows that $\Gamma$ varies linearly with
  $\sqrt{E}$. This agrees with the Wigner threshold law. As for the
  energy dependence of $\Delta_{\text{fesh}}$, it can be assumed to be
  constant and approximated by its value at zero energy.}
\label{gammafit}
\end{figure}

To specify our model, we need to find the parameters that characterize
the Feshbach resonance, $\Gamma$ and $\Delta_{\text{fesh}}$. This is
most easily done by fitting the data to the derivative of the phase
shift $\delta(E,B)$ with respect to $B$, as determined by
Eq.~(\ref{feshphase}),
\begin{equation}
\frac{\partial}{\partial B}\delta(E,B)=
\frac{E'_b(B) \Gamma(E)/2}{(E-E_b(B)-\Delta_{\rm{fesh}})^2+(\Gamma(E)/2)^2}.
\label{deriv}
\end{equation}
The resonance width and the location of the dressed bound state can be
obtained by fitting a Lorentzian to the numerically calculated values
as a function of $E$. In Fig.~\ref{gammafit} we plot the values
obtained from this procedure. The fit to the data agrees well with the
expected Wigner threshold law, $\Gamma/2=C\sqrt{E}$, where we find
$C=2.49$ $\sqrt{\mu K}$. We neglect the small value of the zero
intercept, typically associated with the presence of the inelastic
component.

Finally, from 
Fig.~\ref{gammafit} we see that for $E<0.2$ $\mu$K, the level shift
acquires a constant value of $\Delta_{\text{fesh}}(0)\approx -0.79$
$\mu$K.  Thus, we model the scattering phase shift by
\begin{equation}
\delta(E)=-a_{\text{bg}}k-\tan^{-1}\left[\frac{C \sqrt{E}}{E-E_b(B)
-\Delta_{\text{fesh}}(0)} \right].
\end{equation}
A more complete model at higher energies requires a determination of
the threshold law for $\Delta_{\text{fesh}}$. Marcelis {\em et al.}
\cite{marcelis} have shown for a one channel model or more (if losses
are neglected) that this law can be obtained from the below-threshold
behaviour of the Feshbach bound state as a function of the $B$
field. While this is true, a direct application of their model is not
possible here since we are dealing with a situation where one of the
closed channel gets opened. As we are mainly interested in
near-threshold behaviour in order to extract information below
threshold, we will neglect the functional behaviour of
$\Delta_{\text{fesh}}$ at higher energies.

\begin{figure}[t]
  \includegraphics[scale=.48]{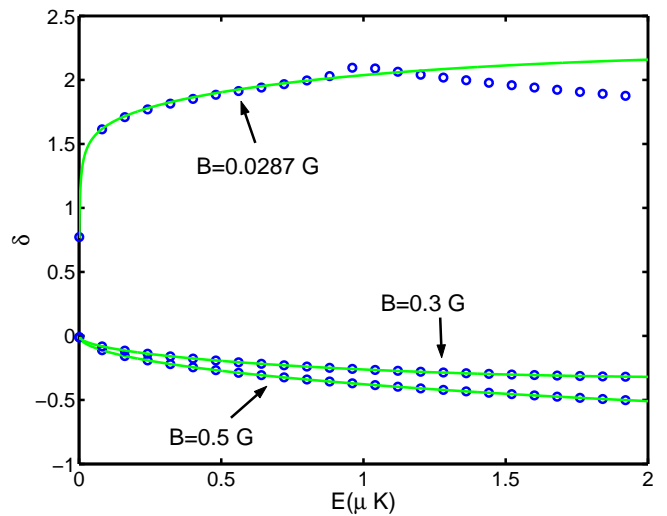}
\caption{ Comparison between the analytical model and the multichannel
  data for scattering on the $\ket{ao}$ channel.  The circles
  represent values of $\delta$ obtained from multichannel code for
  $B=0.0287$, $0.3$, and $0.5$ Gauss. The solid lines correspond to
  the analytical model developed in this paper. The fit is chosen to
  be made in the regime where the |bp> channel is closed.}
\label{phasefitfeshbach}
\end{figure}

\begin{figure}[b]
  \includegraphics[scale=.48]{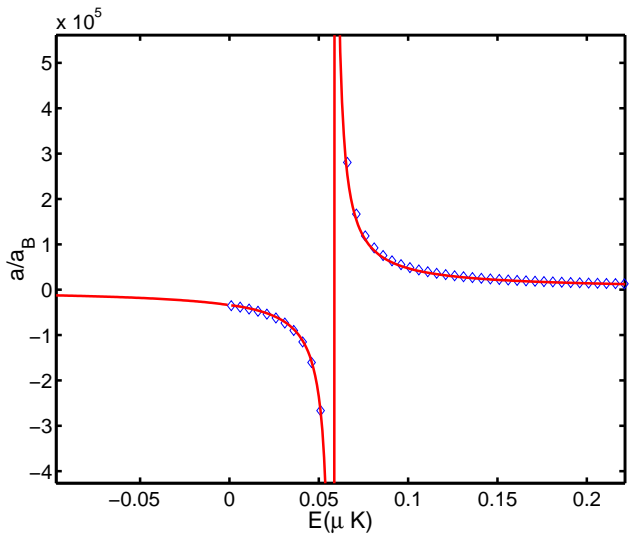}
\caption{Figure shows the analytical continuation of the scattering
  length for $\ket{ao}$ channel to negative energies for $B=0.0287$
  Gauss. There is very good agreement between values obtained using
  the coupled channel code (shown by diamonds) and the analytic
  model (solid line).}
\label{scattfit}
\end{figure}

As in the case of $\ket{ap}$ channel, the background scattering length
acts as a free parameter that is determined by fitting the above model
to one low-energy data point. In Fig.~\ref{phasefitfeshbach} we plot
the scattering phase shift $\delta(B,E)$ for three different values of
the magnetic field $B$. The plots show a very good agreement of the
analytical model to the numerical multichannel data.

Given our model, it is simple to obtain the scattering properties such
as the $s$-wave scattering length at negative energy (below
threshold), by analytically continuing the above formula to the
positive imaginary axis in the complex $k$-plane. Figure
\ref{scattfit} shows excellent agreement of the scattering lengths
obtained using our model with the numerical data for positive
energies. The predicted values at negative energies are also shown
which agree well with the numerical value obtained just above
threshold.

\begin{figure}[t]
  \includegraphics[scale=.48]{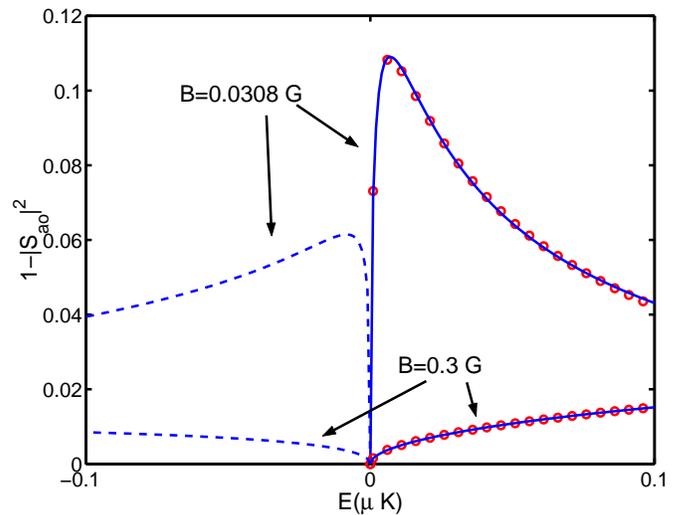}
\caption{Inelastic transitions, $1-|S_{\text{ao}}|^2$, plotted as a
  function of energy. A fit of the analytic inelastic model
  Eq.~(\ref{iemodel}) is shown by solid line to the coupled channel
  numerical shown by red circles. The dashed line corresponds to the
  prediction for negative energies based on analytic continuation of
  the model. }
\label{lossmodel}
\end{figure}

Critical to the use of the TIR for quantum coherent coupling between
atoms is a favorable ratio of elastic to inelastic scattering
processes.  Near threshold scattering at energies $E<E_T(B)$ can
couple the $\ket{ao}$ channel to one other open channel, $\ket{aa}$.
Such transitions will lead to loss of the atomic pair, as the
hyperfine energy will be converted to kinetic energies that are much
larger than the depth of the trap.  By unitarity of the $S$-matrix,
such inelastic processes imply $|S_{ao}| <1$, which can be modeled by
an imaginary part of the scattering phase shift.  Formal theory of
multichannel threshold scattering allows us to model the energy
dependence of this imaginary phase shift
\cite{wigner,ross,jochain}. Nesbet has shown how to separate the
threshold behaviour arising from smooth background and singularities
of poles such as those resulting from virtual bound states and
Feshbach resonances below threshold \cite{nesbet}. For a general
multichannel situation consisting of $M$ open and $n$ closed channels
below and above a certain threshold, the off-diagonal elements of the
$S$-matrix near threshold can be written as
\begin{equation}
S_{i,M+j}=e^{i(\delta_{\text{bg,r}}+i\delta_{\text{bg,r}})}
\gamma_{i,M+j}\frac{2\kappa_{\text{pole}}
  k^{1/2}}{k+i\kappa_{\text{pole}}},
\end{equation} 
where $\gamma$ is some $M\times n$ matrix, $i\in \{1,M\}$, $j\in
\{1,n\}$, and $\kappa_{\text{pole}}$ is the location of the pole on
the imaginary $k$ axis. In our case, for $B>0$, near the $\ket{ao}$
channel threshold, this corresponds to one open channel $\ket{aa}$
and one closed channel $\ket{ao}$, the relevant S-matrix element has
the form
\begin{equation}
|S_{12}|^2=1-|\bra{ao}S \ket{ao}|^2=A_{\text{ie}}e^{-2\delta_{\text{bg,i}}}
\frac{k}{k^2+\kappa_{\text{pole}}^2}, \label{iemodel}
\end{equation}
where $A_{\text{ie}}$ is some proportionality constant that can be
treated as a parameter. The imaginary part of the background phase
shift is given by the Wigner threshold law $\delta_{\text{bg,i}} =
-a_{\text{bg,i}}k$ where $a_{\text{bg,i}}$ is the imaginary part of
the background scattering length.

The values of the parameters of the model are typically obtained by
fitting Eq.~(\ref{iemodel}) to the numerical coupled channels data
once the position of the pole, $\kappa_{\text{pole}}$, is known.
Marcelis {\em et al.} \cite{marcelis} showed that the existence of a
virtual bound state leads to a rapid change in the shift in the energy
of a Feshbach resonance, $\Delta_{\rm{fesh}}$ in Eq.~(\ref{deriv}),
and derived a simple formula that connects the pole location to the
shift.  From inset of Fig.~\ref{boundstate}, the rapid change in
$\Delta_{\rm{fesh}}$ near threshold strongly indicates the existence
of a such a virtual pole.  We cannot, however, apply the formula in
\cite{marcelis}, since that analysis was limited to a two-channel
model.  Here, we have a more complicated interplay of three channels,
($\ket{ao}$, $\ket{aa}$, $\ket{bp}$), since the Feshbach resonance
arises from the bound state in the $\ket{bp}$ channel.  A more
detailed analysis is required to determine the location of
$\kappa_{\text{pole}}$ from first principles.  Instead, since
sufficient data points are available from the coupled channels
numerical solution, we treat $\kappa_{\text{pole}}$ as an additional
fitting parameter.

\begin{figure}[t]
  \includegraphics[scale=.48]{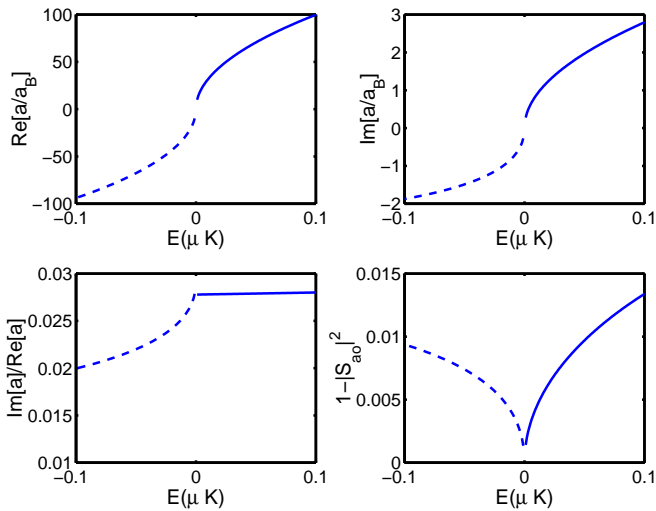}
\caption{Real and imaginary parts of the scattering length  for $\ket{ao}$
 obtained by
  analytic continuation of the multichannel $S$-matrix at $B=0.5$
  G. Solid lines correspond to numerical coupled channels data while
  dashed line is the analytic continuation. }
\label{realandimaglength}
\end{figure}

Figure \ref{lossmodel}
shows excellent fit of the above model to the numerical data. Analytic
continuation of this model will give a precise determination of the
off-diagonal elements of the $S$-matrix at negative energies. 
This is shown in Fig.~\ref{realandimaglength} for the magnetic field
value $B = 0.5$ G. Since the above model for the inelastic component of
the $S$-matrix is known to be valid only in a small energy region
close to threshold, we have limited ourselves to the energy range
$-0.1\mu K$ to $0.1\mu K$ for purpose of illustration. 

As a measure of the ratio of inelastic to elastic processes, we define
a general energy-dependent complex scattering length,
\begin{equation}
\text{Re}[a]+i \text{Im}[a] = -\tan[\delta(E)]/k,
\end{equation}
which can be analytically continued to negative energies. In
Fig.~\ref{realandimaglength}.  we plot $\text{Im}[a]/\text{Re}[a]$.
We see that inelastic processes should not dominate, even at negative
energies sufficiently far from the Feshbach resonance.

\section{Summary and Conclusion}
In this paper we have discussed the need for an analytic model of the
$S$-matrix in order to extract scattering properties at negative
energies where numerical methods fail.  Such negative energy solutions
are essential for understanding trap-induced resonances that involve
atoms tunneling into regions of the molecular potential that are below
the threshold.  Beyond improved numerical solutions, these models give
us physical intuition with regards to the scattering resonances that
are critical to developing many-body model Hamiltonians that can
help to explain ultra-cold atomic phenomena \cite{kokkelmansandholland,
  servaas}.

We applied this method to study the collision of two $^{133}$Cs atoms
in separate harmonic traps, a situation similar to that of atoms in
separate sites of a polarization gradient optical lattice.  Colliding
atoms in spin states that are chosen so that they are robust with
respect to trapping inhomogeneities are not necessarily optimal when
considering the scattering process; they can undergo
multichannel scattering processes, including inelastic loss.  We
studied a specific example of such scattering -- collisions between
$\ket{F=3, M=3}$ and $\ket{F=4, M=3}$.  A Feshbach resonance occurs at
very small magnetic fields ($\sim$ 30 mG) due to coupling to a bound
state in the $\ket{F=3, M=2}\ket{F=4, M=4}$ channel.  To treat this,
we modeled the $S$-matrix element via a smooth background component
with an imaginary part in its scattering length and elastic
single-resonance Feshbach model.  The resulting total scattering
length agrees well with direct numerical multichannel scattering
solution at small positive energies and extends analytically to
negative energies well beyond the validity of numerical solutions.

We have also discussed a model consisting of a few parameters in order
to describe the off-diagonal element of $S$-matrix that corresponds to
inelastic processes. The model agrees extremely well with the coupled
channels calculation at positive energies above threshold. Analytic
continuation allows us to calculate the scattering length at
negative energies below threshold.  We note that, whereas for free
space scattering the reaction rates for elastic and inelastic
processes just above threshold are simply related to the complex
scattering length \cite{BohnJulienne}, in the negative energy case one
must account for the tunneling rate from trap to the molecular
potential.  We will treat this in detail in a future publication.

Within the framework described in this paper, it is also possible to
study the threshold behaviour arising due to the opening/closing of
the $\ket{bp}$ channel. This particular threshold is interesting
because the channel closing can be controlled by use of a very small
magnetic field.  Also since $\ket{bp}$ is the same channel that has
the Feshbach bound state, opening of this channel leads to the
disappearance of this Feshbach resonance. While Feshbach resonances in
ultra-cold atomic gases have been thoroughly studied in recent years,
our study opens up the prospect of studying Feshbach resonances in the
vicinity of such tunable thresholds and their implications to the
many-body properties of trapped ultra-cold gases.

We are extremely grateful to Paul Julienne for his hospitality during our visit 
to NIST in relation to this project and for his direction in the operation of the 
Mies-Julienne-Sando NIST close-coupling codes, used to perform the multichannel 
scattering calculations presented here.  We also thank Rene Stock for helpful discussions
on the trap-induced resonances.  SB and IHD acknowledge financial support from the ONR, Grant
No. N00014-03-1-0508, and DTO Grant No. DAAD19-13-R-0011. SB also acknowledges financial support from
the W. M. Keck Program in Quantum Materials at Rice University. SK acknowledges financial support from the NWO.

\end{document}